\documentclass[preprint,showpacs,amsmath,amssymb,prl,superscriptaddress]{revtex4}
\usepackage{graphicx}% Include figure files
\usepackage{dcolumn}% Align table columns on decimal point
\usepackage{bm}% bold math
\begin{document}

%\preprint{APS/123-QED}

\title{Controlling ultracold Rydberg atoms in the quantum regime}
\date{\today}

\author{Bernd Hezel}
\email[]{hezel@physi.uni-heidelberg.de}
\affiliation{%
Physikalisches Institut, Universit\"at Heidelberg, Philosophenweg
12, 69120 Heidelberg, Germany}
\author{Igor Lesanovsky}
\email[]{igor@iesl.forth.gr}
\affiliation{%
Institute of Electronic Structure and Laser,
Foundation for Research and Technology - Hellas, P.O. Box 1527, GR-711 10 Heraklion, Greece}%
\author{Peter Schmelcher}
\email[]{Peter.Schmelcher@pci.uni-heidelberg.de}
\affiliation{%
Physikalisches Institut, Universit\"at Heidelberg, Philosophenweg 12, 69120 Heidelberg, Germany}%
\affiliation{%
Theoretische Chemie, Institut f\"ur Physikalische Chemie,
Universit\"at Heidelberg,
INF 229, 69120 Heidelberg, Germany}%

\date{\today}

\begin{abstract}\label{txt:abstract}
We discuss the properties of Rydberg atoms in a magnetic
Ioffe-Pritchard trap being commonly used in ultracold atomic
physics experiments. The Hamiltonian is derived and it is
demonstrated how tight traps alter the coupling of the atom to the
magnetic field. We solve the underlying Schr\"odinger equation of
the system within a given $n$-manifold and show that for a
sufficiently large Ioffe field strength the $2n^2$-dimensional
system of coupled Schr\"odinger equations decays into several
decoupled multicomponent equations governing the center of mass
motion. An analysis of the fully quantized center of mass and
electronic states is undertaken. In particular, we discuss the
situation of tight center of mass confinement outlining the
procedure to generate a low-dimensional ultracold Rydberg gas.
\end{abstract}

\maketitle Rydberg atoms possess remarkable properties. Although
being electronically highly excited they can possess lifetimes
of the order of milliseconds. Moreover, due to the large
displacement of the valence electron and the atomic core they
are highly susceptible to electric fields and therefore easily
polarizable. The latter is impressively shown in ultracold Rydberg
gases \cite{Mourachko98} where the mutual multipole interaction of
two Rydberg atoms leads to a number of intriguing many body effects, such as
resonant F\"orster transitions \cite{Gallagher82} and the dipole
blockade mechanism \cite{Lukin01,Tong04}. Their tunable electric
dipole moment makes them also interesting candidates for the
realization of a two-qubit quantum gate \cite{Jaksch00} which is a
crucial ingredient for the realization of quantum information
processing schemes. However, a prerequisite for realizing the
latter in a coherent fashion is the availability of a well-controlled
environment in which the Rydberg-Rydberg interaction takes
place: Firstly one has to assure the Rydberg atoms to be
individually addressable. Secondly, since the mutual interaction
strength strongly depends on their relative displacement, a
suitable way to spatially arrange the atoms has to be found.

There exist several proposals for building traps for Rydberg atoms
suggesting the use of electric \cite{Hyafil04}, optical
\cite{Dutta00} and magnetic fields \cite{Lesanovsky05_03}. The
experimental feasibility of magnetic trapping has been shown
recently by employing very strong magnetic fields \cite{Choi05_1}.
Trapping in the quantum regime, which is indispensable to gain
precise control over the atomic motion, however, could not yet be
demonstrated. Properties of such quantized Rydberg states that
emerge in a magnetic quadrupole trap have been theoretically
discussed in Refs.~\cite{Lesanovsky05_03}. The authors report on a
strong confinement of the atomic center of mass (c.m.) with
trapping frequencies of the order of 10 kHz. At the same time the
electronic structure is, to a large extent, unchanged compared to
the field-free case.  Although the 3D magnetic quadrupole field is
omnipresent in ultracold atom experiments it has one drawback. Due
to the point of zero field strength at its center Majorana
transitions are likely to happen thereby leading to loss from the
trap. In Ref. \cite{Lesanovsky05_03} it was shown that by
increasing the {\it{total angular momentum}} of the atom this
problem can be overcome. However, for practical purposes it is
desirable to have trapping at small or even zero center of mass
angular momentum since this is the regime in which trapped ground
state atoms are usually prepared.

In this letter we show that trapped and controllable Rydberg
states can be achieved in a Ioffe-Pritchard (IP) trap. We discuss
how the large size of the Rydberg atom modifies the coupling to
the magnetic field in comparison to ground state atoms and
demonstrate the feasibility of generating low dimensional Rydberg
gases. Using the IP configuration is of particular relevance since
it is usually employed for preparing ultracold atomic gases which
can serve as the initial state for the production of trapped
ultracold Rydberg atoms.

In the two-body picture the mutual interaction of the valence
electron (particle $1$) and the remaining core (particle $2$) of
the Rydberg atom can be modeled by an effective potential that
exhibits short range properties describing the electron-core
scattering and a long range Coulombic behaviour. States with large
{\it{electronic angular momenta}} $l$, which represent the focus
of the present investigation, probe almost exclusively the
Coulombic tail of this potential. We do not account for
relativistic effects such as spin-orbit coupling which for large
$n,l$ are negligibly small compared to the energy shift due to the
magnetic field of the IP trap \cite{Bethe77,Lesanovsky05_03}. The
interaction of the electronic and core charge with the magnetic
field are taken into account via the minimal coupling. Following
the above arguments and including the coupling of the nuclear and
electronic magnetic moments to the external field our initial
Hamiltonian reads
\begin{eqnarray}
  H_\text{init}&=&\frac{1}{2M_1}\left[\mathbf{p}_1-e\mathbf{A}(\mathbf{r}_1)\right]^2+\frac{1}{2M_2}\left[\mathbf{p}_2+e\mathbf{A}(\mathbf{r}_2)\right]^2 \nonumber\\
  &&+V(\left|\mathbf{r}_1-\mathbf{r}_2\right|)
 -\mbox{\boldmath$\mu$}_1\cdot\mathbf{B}(\mathbf{r}_1)-\mbox{\boldmath$\mu$}_2\cdot\mathbf{B}(\mathbf{r}_2).
  \label{eq:single_particle_hamiltonian_approx}
\end{eqnarray}
The IP field configuration is given by
$\mathbf{B}(\mathbf{r})=B\mathbf{e}_z+\mathbf{B}_\text{lin}(\mathbf{r})+\mathbf{B}_\text{quad}(\mathbf{r})$
with the linear component
$\mathbf{B}_\text{lin}(\mathbf{r})=G\left[x\mathbf{e}_x-y\mathbf{e}_y\right]$
and the quadratic component
$\mathbf{B}_\text{quad}(\mathbf{r})=Q\left[2z(x\mathbf{e}_x+y\mathbf{e}_y)+(x^2+y^2-2z^2)\mathbf{e}_z)\right]$.
In the following we consider the case of a dominant linear
component and neglect $\mathbf{B}_\text{quad}(\mathbf{r})$. We
therefore encounter a two-dimensional quadrupole field which is
translationally invariant along the $z-$axis. Our corresponding
vector potential reads
$\mathbf{A}(\mathbf{r})=\mathbf{A}_\text{c}(\mathbf{r})+\mathbf{A}_\text{lin}(\mathbf{r})$
with
$\mathbf{A}_\text{c}(\mathbf{r})=\frac{B}{2}\left[x\mathbf{e}_y-y\mathbf{e}_x\right]$
and $\mathbf{A}_\text{lin}(\mathbf{r})=Gxy\mathbf{e}_z$. Next let
us insert these expressions into the Hamiltonian
(\ref{eq:single_particle_hamiltonian_approx}) and introduce
relative $\mathbf{r}$ and c.m.~coordinates $\mathbf{R}$ with their
respective momenta $\mathbf{p}$ and $\mathbf{P}$. In the absence
of the external field one arrives at a decoupled c.m.~and
electronic motion. However, already in the presence of a
homogeneous field this is not the case \cite{Schmelcher92_94} and
consequently terms coupling the c.m.~and internal motion emerge.
To simplify the latter in case of our inhomogeneous magnetic field
configuration we apply the unitary transformation
$U=\exp\left[i\frac{B}{2}\mathbf{e}_z\times \mathbf{r} \cdot
\mathbf{R}\right]$ which eliminates couplings of the c.m. and the
relative motion generated by the Ioffe field (atomic units are
used throughout except when stated otherwise) \footnote{$\hbar=1$,
$M_1=m_e=1$, $a_0=1$: The magnetic gradient unit then becomes
$b=1a.u.=4.44181\cdot 10^{15} \frac{T}{m}$. The magnetic field
strength unit is $B=1a.u.=2.35051\cdot 10^{5} T$}. Furthermore we
neglect the diamagnetic terms, which is a very good approximation
for laboratory field strengths and not too high principal quantum
numbers $n$ (see ref. \cite{Lesanovsky05_03}), and keep only the
leading order terms with respect to the inverse masses
\begin{eqnarray}
  H_\text{IP}=H_A+\frac{\mathbf{\mathbf{P}}^2}{2M_2}
  -\mbox{\boldmath$\mu$}_2\cdot \mathbf{B}(\mathbf{R})
    -\mbox{\boldmath$\mu$}_1\cdot \mathbf{B}(\mathbf{R}+
\mathbf{r})+\mathbf{A}_\text{lin}(\mathbf{R}+\mathbf{r})\cdot\mathbf{p}. \label{eq:hamiltonian_approximated}
\end{eqnarray}
Here $H_A=\frac{\mathbf{p}^2}{2}-\frac{1}{r}$ is the Hamiltonian
of a hydrogen atom possessing the eigenfunctions
$\left|n,l,m_l,m_s\right>$ and energies
$E_{n}=-\frac{1}{2}n^{-2}$. The following two terms of
$H_\text{IP}$ describe the c.m.~motion of a point particle
possessing the magnetic moment $\mbox{\boldmath$\mu$}_2$ in the
presence of the field $\mathbf{B}$. This system has been
thoroughly investigated in Refs. \cite{Lesanovsky04_4,Bill06}.
Since the magnetic moments are connected to the corresponding
spins $\mathbf{S}$ and $\mathbf{\Sigma}$ according to
$\mbox{\boldmath$\mu$}_1=-\mathbf{S}$ and
$\mbox{\boldmath$\mu$}_2=-\frac{g_N}{2M_2}\mathbf{\Sigma}$, with
$g_N$ being the nuclear $g$-factor, we neglect the term involving
$\mbox{\boldmath$\mu$}_2$ in the following. The last two terms of
$H_\text{IP}$ couple the electronic and c.m.~dynamics mediated by
a spin-field and motionally induced coupling. We remark that the
Hamiltonian (\ref{eq:single_particle_hamiltonian_approx}) commutes
with the $z$-component of the linear c.m.~momentum $P_z$ being a
direct consequence of the above-mentioned translational invariance
of the system along the $z$-axis. Hence the longitudinal motion
can be integrated out employing plane waves
$\left|K_z\right>=\exp\left(-i K_z Z\right)$. In order to solve
the eigenvalue problem of the resulting Hamiltonian, that depends
on five spatial degrees of freedom, we assume the magnetic field
not to couple adjacent $n$-manifolds. Estimating the energy level
shift caused by the magnetic field according to
$E_\text{zee}\approx B n$ this requirement is fulfilled if
$|E_{n}-E_{n-1}|/E_\text{zee}\approx \left(B n^4\right)^{-1} \gg
1$. In this case we can consider each $n$-manifold separately and
represent the Hamiltonian (\ref{eq:hamiltonian_approximated}) in
the space of the $2n^2$ states spanning the given $n$-manifold.
Neglecting the constant energy offset $E_n$ and introducing scaled
c.m.~coordinates ($\mathbf{R}\rightarrow \gamma^{-\frac{1}{3}}
\mathbf{R}$ with $\gamma=G M_2$) while scaling the energy unit via
$\epsilon_\text{scale}=\gamma^\frac{2}{3}/M_2$ we eventually
arrive at the working Hamiltonian
\begin{eqnarray}
  \mathcal{H}&=&\frac{P_x^2+P_y^2}{2}+\mbox{\boldmath$\mu$}\cdot
  \mathbf{G}(X,Y)+\gamma^{-\frac{2}{3}}M_2\mathcal{H}_r\label{eq:working_hamiltonian}
\end{eqnarray}
with the effective magnetic field
\begin{eqnarray}
\mathbf{G}(X,Y)&=&X\mathbf{e}_x-Y\mathbf{e}_y+\gamma^{-\frac{2}{3}}M_2B\mathbf{e}_z
\end{eqnarray}
where $\mbox{\boldmath$\mu$}$ and $\mathcal{H}_r$ are the
$2n^2$-dimensional matrix representations of the operators
$1/2\left[\mathbf{L}+2\mathbf{S}\right]$ and
$H_r=\mathbf{A}_\text{lin}(\mathbf{r})\cdot\mathbf{p}+\mathbf{B}_\text{lin}(\mathbf{r})\cdot\mathbf{S}$,
respectively. Here we have introduced the orbital angular momentum
operator $\mathbf{L}=\mathbf{r}\times \mathbf{p}$.

The Hamiltonian (\ref{eq:working_hamiltonian}) can be interpreted
as follows: The first two terms describe the dynamics of a neutral
atom in a IP trap under the assumption that the coupling of the
atom to the field is given by the Zeeman energy
$E_z=1/2\left[\mathbf{L}+2\mathbf{S}\right]\cdot \mathbf{G}(X,Y)$.
One encounters a similar coupling term also for 'point-like'
ground state atoms \cite{Bill06} where, however, the generically
strong hyperfine coupling leads to the fact that
$\mbox{\boldmath$\mu$}$ is proportional to the total spin
$\mathbf{F}$. The distinct feature of the Hamiltonian
(\ref{eq:working_hamiltonian}) is the appearance of the last term
which accounts for the finite size of the Rydberg atom. This term
scales according to $\gamma^{-\frac{2}{3}}GM_2
n^2=\gamma^{\frac{1}{3}}n^2\approx
\gamma^{\frac{1}{3}}\left<r\right>$. Since $\gamma^{-\frac{1}{3}}$
can be regarded as a length unit for the c.m.~wave function we
find this term to be particularly important if
$\left<r\right>\approx\left<R\right>$, i.e. if the size of the
atom and the size of the c.m.~state become comparable. In a
typical macroscopic IP trap \cite{Choi05_1} the c.m.~wave
functions are very extended and the energy spacing between the
c.m.~states is small. Hence, in this 'classical' regime
$\mathcal{H}_r$ has little effect and can be neglected. The
situation changes for tighter IP traps which can, for example, be
realized by so-called atom chips \cite{Folman02}. Here the micro
structured wires allow for the generation of traps for which the
size of the c.m.~ground states are of the order of several 100 nm
and become therefore comparable to the typical size of Rydberg
atoms.

In order to solve the Schr\"odinger equation belonging to the
Hamiltonian (\ref{eq:working_hamiltonian}) we employ an adiabatic
separation of the electronic and the c.m.~motion. To this end a
unitary transformation $U(X,Y)$ which diagonalizes the last two
(matrix) terms of the Hamiltonian, i.e.
$U^\dagger(X,Y)(\mbox{\boldmath$\mu$}\cdot\mathbf{G}(X,Y)+
\gamma^{-\frac{2}{3}}M_2\mathcal{H}_r)U(X,Y)=E_\alpha(X,Y)$, is
applied. Since $U(X,Y)$ depends on the c.m.~coordinates the
transformed kinetic term involves non-adiabatic (off-diagonal)
coupling terms which we will neglect in the following. We are
thereby lead to a set of $2n^2$ decoupled differential equations
governing the adiabatic c.m.~motion within the individual
two-dimensional energy surfaces $E_\alpha$, i.e. the surfaces
$E_\alpha$ serve as potentials for the c.m. of the atom.
\begin{figure}[htb]\center
\includegraphics[angle=0,width=7cm]{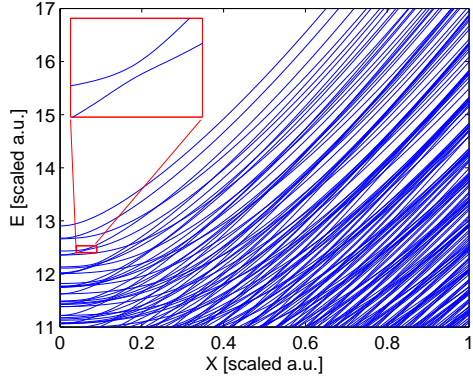}
\caption{Cut through the adiabatic potential surfaces of the
$n=30$-manifold ($^\text{87}\text{Rb}$, $G=20$~T/m, $B=0.01$~G).
The high density of states is clearly visible. The uppermost
potential surface is predominantly formed by the atomic state with
largest angular momentum. This particular surface is clearly
separated from the next lower ones whereas all other surfaces
exhibit a number of avoided crossings (see the magnified view in
the inset) at which non-adiabatic inter-surface transitions are
likely to occur.} \label{fig:cut}
\end{figure}
Figure \ref{fig:cut} shows an intersection through a subset of
these surfaces for the case of $^\text{87}\text{Rb}$ in a IP trap
with a gradient $G=20$~T/m and a Ioffe field strength $B=0.01$~G.
One immediately notices the large number of avoided crossings
between the adiabatic potential surfaces. Here non-adiabatic
transitions mediated by the (neglected) off-diagonal coupling
terms of the kinetic energy are likely to occur. The uppermost
surface, however, does not exhibit such avoided crossings and is a
possible candidate in order to achieve stable trapping. According
to our findings this surface is predominantly formed by the
electronic state possessing the largest possible orbital angular
momentum in the $n$-manifold under consideration. The
corresponding quantum defects and relativistic corrections are
therefore tiny which \textit{a posteriori} justifies their
neglect.

\begin{figure}[htb]\center
\includegraphics[angle=0,width=7.5cm]{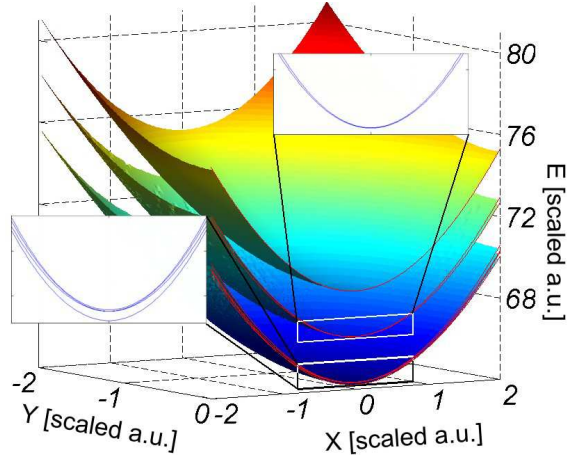}
\caption{Surface plot of the the seven uppermost energy surfaces
of the $n=30$-manifold ($^\text{87}\text{Rb}$, $G=20\,T/m$
$B=0.1\, G$). Clearly the grouping into three multiplets whose
mutual distance is given by $\gamma^{-\frac{2}{3}}M_2B$ is
visible. While the uppermost manifold consists only of a single
surface the next lower ones show an approximate twofold and
fourfold degeneracy. A magnified view of them is provided in the
insets.} \label{fig:3dsurf}
\end{figure}

The actual appearance of the potential surfaces depends on both,
the gradient and the strength of the homogeneous Ioffe field. In
figure \ref{fig:cut} the latter is comparatively small such that
for small displacements from the coordinate center its field
strength is easily surpassed by that of the gradient field. The
different symmetry properties of the both fields lead to a
rearrangement of the energy surfaces finding its expression in the
large number of avoided crossings. We now turn to the case where
the Ioffe field is large compared to that of the gradient field,
i.e. $\gamma^{-\frac{2}{3}}M_2B\gg 1$. This however can hold true
only in the vicinity of the trap center. We assume this region to
be sufficiently large such that at least a few low-lying
c.m.~states are localized here. In this case the term
$\mbox{\boldmath$\mu$}\cdot\mathbf{G}(X,Y)$ will dominate the
Hamiltonian (\ref{eq:working_hamiltonian}). We now diagonalize
this term by employing the transformation $
U_D(X,Y)=\exp\left[i(L_z+
S_z)\phi\right]\exp\left[i(L_y+S_y)\beta\right]$ with
$\phi=\arctan\left[Y/X\right]$,
$\cos\beta=\gamma^{-\frac{2}{3}}M_2B|\mathbf{G}(X,Y)|^{-1}$ and
$\sin\beta=-(X^2+Y^2)|\mathbf{G}(X,Y)|^{-1}$. This yields the
adiabatic energy surfaces
\begin{eqnarray}
  E_\alpha=U^\dagger_D(X,Y)\mbox{\boldmath$\mu$}\cdot\mathbf{G}(X,Y)U_D(X,Y)
  =\frac{1}{2}(L_z+2S_z)|\mathbf{G}(X,Y)|,
\end{eqnarray}
being characterized by the quantum numbers of $L_z$ and $S_z$
which are $m_l$ and $m_s$, respectively. The energetically highest
surface is assumed for $m_l=n-1$ and $m_s=1/2$. The next lower one
is twofold degenerate and the following one shows a fourfold
degeneracy (see figure \ref{fig:3dsurf}). The energy gap between
these degenerate multiplets is given by $\triangle E
\approx\gamma^{-\frac{2}{3}}M_2 B$ and can hence be continuously
varied by tuning the Ioffe field strength. In the present regime
the term $\gamma^{-\frac{2}{3}}M_2 \mathcal{H}_r$ can be
considered as a perturbation since its energetic contribution is
much smaller than $\triangle E$. The correction to the uppermost
surface is zero whereas $\mathcal{H}_r$ couples the surfaces of
the energetically lower lying degenerate multiplets. To study the
dynamics of the multi-component c.m.~wave function within these
coupled potential surfaces constitutes a very interesting problem.
%It is reminiscent of the rovibronic interactions taking place in
%non-Born-Oppenheimer molecular systems which leads to a wealth of
%intriguing phenomena such as local symmetry breaking and ultrafast
%nonadiabatic decay processes \cite{Dom04}.
In the present investigation, however, we will focus exclusively
on the uppermost non-degenerate surface.

The explicit knowledge of $U_D(X,Y)$ allows for an analytical
calculation of the non-adiabatic couplings between any of the
potential surfaces arising from the kinetic energy term. Our
findings show them to be proportional to
$\gamma^{\frac{2}{3}}(M_2B)^{-1}=(\triangle E)^{-1}$. For a
sufficiently large Ioffe field strength we can thus safely employ
the adiabatic approximation, i.e. neglect the non-adiabatic
coupling between the uppermost and the next lower surface.
\begin{figure}[htb]\center
\includegraphics[angle=0,width=6.0cm]{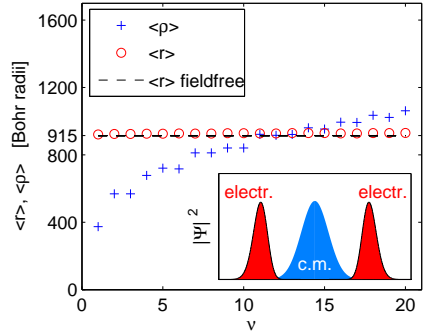}
\caption{Expectation value of the radii $\rho=\sqrt{X^2+Y^2}$ and
$r$ of the c.m.~and electronic wave function in the IP trap
($^\text{87}\text{Rb}$, $G=100\,T/m$, $B=0.1\, G$). $\nu$ labels
the c.m.~quantum states within the uppermost adiabatic potential
surface. While $\left<\rho\right>$ is increasing $\left<r\right>$
remains approximately constant at its field free value. For small
degrees of c.m.~excitation the c.m.~state is even stronger
localized than the valence electron
($\left<\rho\right><\left<r\right>$). Moreover, since the electron
is found in a high angular momentum state its radial uncertainty
$\triangle r$ is small. Thus a scenario where the c.m.~and the
electronic wave function do not overlap is possible as sketched in
the inset.} \label{fig:size}
\end{figure}
In order to obtain the quantized c.m.~states we solve the scalar
Schr\"odinger equation
$\left[1/2(P_x^2+P_y^2)+E_\alpha(X,Y)-\epsilon_\nu\right]\Psi_\nu(X,Y)=0$
in the uppermost potential surface which we denote by $\alpha=2
n^2$. For low c.m.~excitations the potential surfaces are
approximately harmonic and thus the energies are in reasonably
good agreement to those of a two-dimensional isotropic harmonic
oscillator with a $n$-dependent frequency
$\omega=G\sqrt{\frac{n}{2BM_2}}$ (in atomic units). Hence, by
choosing high gradients and an appropriate tuning of the Ioffe
field strength $B$ very tightly confining traps for highly excited
atoms can be obtained. Such a situation is depicted in figure
\ref{fig:size} where we show the expectation values of the radii
$\rho=\sqrt{X^2+Y^2}$ and $r$ for the c.m.~and electronic wave
function versus the degree of excitation of the c.m.~motion $\nu$.
In the presented case the confinement gives rise to a trap
frequency of approximately $1.4\, \text{MHz}$. In this regime the
size of the c.m.~state characterized by $\left<\rho\right>$ is
even smaller than the electronic cloud, i.e. the c.m.~wave
function is stronger localized than the valence electron. On the
other hand the expectation value $\left<r\right>$ for the electron
remains nearly constant possessing the corresponding field free
value as the degree of excitation of the c.m.~increases. This
indicates that in spite of the strong localization of the c.m.~the
electronic structure of the atom is barely changed compared to the
field free case. This observation has been backed up by
calculating further electronic properties, such as the expectation
values of $\mathbf{L}^2$ which also barely differ from their
corresponding field free values. As previously indicated we find
the electron in the highest angular momentum state ($l=n-1$) which
possesses the smallest radial uncertainty $\triangle r$ for given
$n$. Due to this fact it is possible that the c.m.~and the
electronic wave function may not even overlap (see inset of figure
\ref{fig:size}). This novel regime opens up the possibility to
control Rydberg atoms in the quantum regime and might pave the way
to study many-body effects in low-dimensional ultracold Rydberg
gases \cite{Carroll04}.

In order to study the latter, ultracold atoms confined in a tight
atom chip trap \cite{Folman02} can be transferred into high-$l$
Rydberg states by imposing suitable optical and radio frequency
fields (see refs. \cite{Chen93,Lutwak97} and refs. therein). Since
the electronic structure is barely affected even for tight c.m.~confinement
the Rydberg atoms keep their well-known properties
such as long radiative lifetimes and electric dipole moments.

I.L. acknowledges support by the European Community and its 6th Community
Frame under the program of scholarships 'Marie Curie'.
P.S.  acknowledges financial support by the Deutsche Forschungsgemeinschaft.

\end{document}